\documentclass{aastex}
\shorttitle{Toroidal equilibria in spherical coordinates}
\shortauthors{Tsui}
\begin{document}
\title{Toroidal equilibria in spherical coordinates}
\author{K.H. Tsui}
\affil{Instituto de F\'{i}sica - Universidade Federal Fluminense,
\\Campus da Praia Vermelha, Av. General Milton Tavares de Souza s/n
\\Gragoat\'{a}, 24.210-346, Niter\'{o}i, Rio de Janeiro, Brasil.}
\email{tsui$@$if.uff.br}
\date{}
\pagestyle{myheadings}
\baselineskip 24pt

\begin{abstract}

The standard Grad-Shafranov equation for
 axisymmetric toroidal plasma equilibrium is customary expressed
 in cylindrical coordinates with toroidal contours,
 and through which benchmark equilibria are solved.
 An alternative approach to cast the Grad-Shafranov equation
 in spherical coordinates is presented.
 This equation, in spherical coordinates,
 is examined for toroidal solutions
 to describe low $\beta$ Solovev
 and high $\beta$ plasma equilibria
 in terms of elementary functions.
  
\end{abstract}

\vspace{2.0cm}
\keywords{Toroidal Equilibria}
\renewcommand{\thesection}{\Roman{section}}

\maketitle
\newpage
\section{Grad-Shafranov Equation}

The Grad-Shafranov equation (1,2) for toroidal plasma equilibria
 is traditionally formulated in a toroidal system
 using cylindrical coordinates to suit the plasma topology.
 This equation describes the poloidal magnetic flux,
 in cylindrical coordinates,
 with source terms that are themselves flux gradients.
 By specifying the source functionals in magnetic flux,
 toroidal equilibria are solved by different methods.
 Because of the cylindrical representation of a toroidal topology,
 the solutions are not easy to visualize and the mathematical
 analytic tools are rather limilted,
 which weaken the communication
 between theorists and experimentalists.
 Here, we propose to solve axisymmetric plasma equilibria
 in standard spherical coordiantes
 by looking for solutions with toroidal topology.
 With the familiar spherical system,
 and the large volume of well-known special functions
 in this coordinate system,
 the solutions in toroidal topology appear in a
 much friendly presentation,
 which improves physical interpretations.
  
We first derive the Grad-Shafranov equation
 in spherical coordinates by considering

\begin{eqnarray}
\label{eqno1}
\vec J\times\vec B-\nabla p\,
 =\,0\,\,\,,
\\
\label{eqno2}
\nabla\times\vec B\,=\,\mu\vec J\,\,\,.
\end{eqnarray}

\noindent With axisymmetry, the magnetic field
 and the current density
 can be represented by two scalar functions
 in standard spherical coordinates

\begin{eqnarray}
\label{eqno3}
\vec B\,
 =\,\nabla P\times\nabla\phi+Q\nabla\phi\,
 =\,{1\over r\sin\theta}
 \{+{1\over r}{\partial P\over\partial\theta},
   -{\partial P\over\partial r},
   +Q\}\,\,\,,
\\
\label{eqno4}
\mu\vec J\,
 =\,{1\over r\sin\theta}
 \{+{1\over r}{\partial Q\over\partial\theta},
   -{\partial Q\over\partial r},
   -{\partial^2 P\over\partial r^2}
   -{1\over r^2}\sin\theta
   {\partial\over\partial\theta}
   ({1\over\sin\theta}{\partial P\over\partial\theta})\}
   \,\,\,.
\end{eqnarray}

\noindent Making use of Eq.~\ref{eqno2}
 to eliminate the current density,
 Eq.~\ref{eqno1} renders three components.
 The $\phi$ component contains only the magnetic force,
 and it is

\begin{mathletters}
\begin{eqnarray}
\label{eqno5a}
{\partial P\over\partial r}
 {\partial Q\over\partial\theta}
 -{\partial P\over\partial\theta}
 {\partial Q\over\partial r}\,
 =\,0\,\,\,,
\\
\label{eqno5b}
Q(r,\theta)\,=\,Q(P(r,\theta))\,\,\,.
\end{eqnarray}
\end{mathletters}

\noindent As for the $\theta$ component, it reads

\begin{equation}
\label{eqno6}
{\partial P\over\partial\theta}
 \{{\partial^2 P\over\partial r^2}
 +{1\over r^2}\sin\theta{\partial\over\partial\theta}
 ({1\over\sin\theta}{\partial P\over\partial\theta})\}
 +Q{\partial Q\over\partial\theta}
 +\mu r^2\sin^2\theta
 {\partial p\over\partial\theta}\,
 =\,0\,\,\,.
\end{equation}

\noindent Since the spatial dependencies of $Q(r,\theta)$
 is through a functional of $P(r,\theta)$,
 we could express the $\theta$ dependency
 in plasma pressure by $P$ to write $p(r,\theta)=p(r,P)$,
 and the above equation becomes

\begin{eqnarray}
\nonumber
\{{\partial^2 P\over\partial r^2}
 +{1\over r^2}\sin\theta{\partial\over\partial\theta}
 ({1\over\sin\theta}{\partial P\over\partial\theta})
 +{\partial Q\over\partial P}Q\}
 +\mu r^2\sin^2\theta
 {\partial p\over\partial P}\,
\\
\label{eqno7}
=\,\{{\partial^2 P\over\partial r^2}
 +{1\over r^2}\sin\theta{\partial\over\partial\theta}
 ({1\over\sin\theta}{\partial P\over\partial\theta})
 +{1\over 2}{\partial Q^2\over\partial P}\}
 +\mu r^2\sin^2\theta
 {\partial p\over\partial P}\,
 =\,0\,\,\,.
\end{eqnarray}

\noindent This equation is the Grad-Shafranov equation (1,2)
 of axisymmetric toroidal plasma equilibrium,
 represented in spherical coordinate system.
 The first three terms represent
 the nonlinear force-free field with $\mu\vec J=K(P)\vec B$,
 where $K(P)={\partial Q/\partial P}$ is a scalar function.
 This can be verified from Eq.~\ref{eqno3} and Eq.~\ref{eqno4}
 when we impose $\mu\vec J=K(P)\vec B$.
 In particular, we would have the linear force-free field
 should we take $Q(P)=aP$ with constant $K(P)=a$.
 The last term is the plasma pressure balance.
 The magnetic function $Q^{2}(P)$ and the pressure function $p(P)$
 are source functions of the Grad-Shafranov equation
 that need to be specified.
 Finally, the $r$ component of Eq.~\ref{eqno1} reads

\begin{eqnarray}
\nonumber
{d\over dr}p(r,P)
 =\,-{1\over\mu}({1\over r\sin\theta})^2
 {\partial P\over\partial r}
 \{{\partial^2 P\over\partial r^2}
 +{1\over r^2}\sin\theta{\partial\over\partial\theta}
 ({1\over\sin\theta}{\partial P\over\partial\theta})
 +{1\over 2}{\partial Q^2\over\partial P}\}\,
\\
\label{eqno8}
 =\,{\partial P\over\partial r}{\partial\over\partial P}p(r,P)
 \,\,\,.
\end{eqnarray}

\noindent The term $dp(r,P)/dr$ on the left side
 refers to the total radial derivative,
 which has an explicit and an implicit part,
 and the right side of this radial equation
 is the magnetic force.
 Making use of the $\theta$ component, Eq.~\ref{eqno7},
 the right side is equal to the implicit part
 of the radial pressure gradient,
 which cancels the same term on the left side,
 the implicit part of the radial derivative.
 The radial component, therefore,
 reads $\partial p/\partial r=0$,
 and the plasma pressure
 is a function of $P$ only,

\begin{equation}
\label{eqno9}
p\,=\,p(P)\,\,\,.
\end{equation}

\newpage
\section{Low $\beta$ Equilibria}

To solve Eq.~\ref{eqno7} analytically, we take the source
 functions as

\begin{mathletters}
\begin{eqnarray}
\label{eqno10a}
Q^2(P)\,=\,2b^2P+a^2P^2+Q^{2}_{0}\,\,\,,
\\
\label{eqno10b}
p(P)\,=\,p_{0}P+p^{'}_{0}\ln P-C_{0}\,\,\,,
\end{eqnarray}
\end{mathletters}

\noindent where $p_{0}$ and $p^{'}_{0}$ carry
 the dimension of pressure,
 $a^2$ and $b^2$ measure the dimension of magnetic field,
 $Q^{2}_{0}$ and $C_{0}$ are constants,
 and $P(r,\theta)$ is a dimensionless function.
 With $a^2=0$ and $p^{'}_{0}=0$, Solovev (3) solved
 the Grad-Shafranov equation for low $\beta$
 equilibria with linear source functions.
 These source functions have been extended
 by other investigators (4,5).
 With $a^2\neq 0$, the quadratic form for $Q^{2}$
 plus a likewise quadratic form for $p$,
 without the $\ln P$ term,
 have been analyzed by many authors in different ways (6-9).
 In particular, Hu has considered a cubic form for $Q^{2}$ (10).
 On the other hand, numerical routines have also been advanced
 by many authors (11,12).
 Here, we include a $\ln P$ term in the plasma pressure function,
 and regard Eq.~\ref{eqno10a} and Eq.~\ref{eqno10b}
 as the generalized Solovev case,
 and solve for the Grad-Shafranov equation
 in spherical coordinates for a low $\beta$ plasma.
 The $\theta$ component, therefore, becomes

\begin{eqnarray}
\label{eqno11}
\{r^{2}{\partial^2 P\over\partial r^2}
 +\sin\theta{\partial\over\partial\theta}
 ({1\over\sin\theta}{\partial P\over\partial\theta})
 +(ar)^{2}P+(br)^2\}
 +\mu p_{0} r^{4}\sin^2\theta
 +\mu p^{'}_{0} r^{4}\sin^2\theta{1\over P}\,
 =\,0\,\,\,.
\end{eqnarray}

\noindent Dividing this equation over by $P$,
 separating the variables by writing
 $P(r,\theta)=R(r)\Theta(\theta)$,
 denoting $x=\cos\theta$,
 and with $n(n+1)$ as the separation constant, we have

\begin{mathletters} 
\begin{eqnarray}
\label{eqno12a}
(1-x^2){d^2\Theta(x)\over dx^2}
 +n(n+1)\Theta(x)\,
 =\,-{(br)^2\over R(r)}
 -{\mu p^{'}_{0}\over b^4}{(br)^4\over R^2(r)}
 {(1-x^2)\over\Theta(x)}\,\,\,,
\\
\label{eqno12b}
r^{2}{d^2R(r)\over dr^2}
 -n(n+1)R(r)\,
 =\,-(ar)^{2}R(r)
 -{\mu p_{0}\over b^4}(br)^{4}
 {(1-x^2)\over\Theta(x)}\,\,\,.
\end{eqnarray}
\end{mathletters}

\noindent These equations can be solved by iterations,
 because of the low $\beta$ plasma.
 We, therefore, first solve the homogeneous equations
 with the right sides null giving

\begin{mathletters}
\begin{eqnarray}
\label{eqno13a}
\Theta_{0}(x)\,=\,(1-x^2){dP_{n}(x)\over dx}\,\,\,,
\\
\label{eqno13b}
R_{0}(r)\,=\,A_{0}(br)^{n+1}\,\,\,,
\end{eqnarray}
\end{mathletters}

\noindent where $P_{n}(x)$ is the Legendre polynomial.
 The coefficient $A_{0}$ reflects the amplitude
 of the magnetic flux through Eq.~\ref{eqno3}.
 To iterate on the zeroth order solutions,
 we take $n=1$ and substitute Eq.~\ref{eqno13a} and
 Eq.~\ref{eqno13b} to the right sides of
 Eq.~\ref{eqno12a} and Eq.~\ref{eqno12b} to get

\begin{mathletters} 
\begin{eqnarray}
\nonumber
(1-x^2){d^2\Theta_{1}(x)\over dx^2}
 +n(n+1)\Theta_{1}(x)\,
 =\,-{(br)^2\over R_{0}(r)}
 -{\mu p^{'}_{0}\over b^4}{(br)^4\over R_{0}^2(r)}
 {(1-x^2)\over\Theta_{0}(x)}\,
\\
\label{eqno14a}
 =\,-\{{1\over A_{0}}
 +{\mu p^{'}_{0}\over b^4}{1\over A^{2}_{0}}\}\,
 =\,-{1\over A^{*}_{0}}\,\,\,,
\\
\nonumber
r^{2}{d^2R_{1}(r)\over dr^2}
 -n(n+1)R_{1}(r)\,
 =\,-(ar)^{2}R_{0}(r)
 -{\mu p_{0}\over b^4}(br)^{4}
 {(1-x^2)\over\Theta_{0}(x)}\,
\\
\label{eqno14b}
 =\,-\{A_{0}({a\over b})^2+{\mu p_{0}\over b^4}\}(br)^4\,
 =\,-A_{1}(br)^4\,\,\,.
\end{eqnarray}
\end{mathletters}

\noindent By simple inspection, the first order solutions
 and the general solutions are

\begin{mathletters}
\begin{eqnarray}
\nonumber
\Theta_{1}(x)\,=\,-{1\over 2A^{*}_{0}}\,\,\,,
\\
\label{eqno15a}
\Theta(x)\,=\,(1-x^2)-{1\over 2A^{*}_{0}}\,\,\,,
\\
\nonumber
R_{1}(r)\,=\,-{A_{1}\over 10}(br)^{4}\,\,\,,
\\
\label{eqno15b}
R(r)\,=\,A_{0}(br)^{2}-{A_{1}\over 10}(br)^{4}\,\,\,.
\end{eqnarray}
\end{mathletters}

\noindent We see that $\Theta(x)$ has a lobed solution
 centered at $x=0$ with a maximum amplitude $(1-1/2A^{*}_{0})$.
 The radial solution $R(r)$ is composed of two terms.
 The first term is positive
 and increases to the second power of $(br)$,
 while the second term is negative
 and increases to the fourth power of $(br)$.
 The radial function $R(r)$ is positive at small $(br)$
 and negative at large $(br)$.
 It vanishes at $(br)^2=0$ and $(br)^2=10A_{0}/A_{1}$,
 with a maximum in between.
 This maximum of $R(r)$ and the lobed solution of $\Theta(x)$
 are essential for a torus structure for the magnetic flux.
 With the given source functions,
 we notice that the iteration scheme
 can not be carried beyond the first round.
 Consequently, if Eq.~\ref{eqno15a} and Eq.~\ref{eqno15b}
 have any meaning at all,
 $R_{1}(r)$ has to be much less than $R_{0}(r)$,
 and $\Theta_{1}(x)$ has to be much less than $\Theta_{0}(x)$,
 leading to $A_{1}/A_{0}<<1$ and $1/A^{*}_{0}<<1$.
 This requires $[(a/b)^2+(\mu p_{0}/b^{4}A_{0})]<<1$,
 or $(a/b)^2<<1$ and $(\mu p_{0}/b^{4}A_{0})<<1$,
 and $[1+(\mu p^{'}_{0}/b^{4}A_{0})]<<A_{0}$,
 for a low $\beta$ plasma.
 
\newpage
\section{Magnetic Torus}

With the spatial structure solved,
 the magnetic field components are given by

\begin{mathletters}
\begin{eqnarray}
\label{eqno16a}
B_{r}\,=\,+{1\over r\sin\theta}
 {1\over r}{\partial P\over\partial\theta}\,
 =\,-{1\over r^2}R(r)
 {d\Theta(x)\over dx}\,\,\,,
\\
\label{eqno16b}
B_{\theta}\,=\,-{1\over r\sin\theta}
 {\partial P\over\partial r}\,
 =\,-{1\over r}{dR(r)\over dr}
 {1\over (1-x^2)^{1/2}}\Theta(x)\,\,\,,
\\
\label{eqno16c}
B_{\phi}\,=\,+{1\over r\sin\theta} Q(P)\,\,\,.
\end{eqnarray}
\end{mathletters}

\noindent The solution $R(r)$ vanishes at some $r$
 where we have $B_{r}(r)=0$.
 The solution $\Theta(x)$ also vanishes at some $x$.
 Together they describe the magnetic fields.
 Within this region of $(r,x)$, the topological center
 defined by $dR(r)/dr=0$ and $d\Theta(x)/dx=0$
 has $B_{r}=0$ and $B_{\theta}=0$.
 This is the magnetic axis, $r=r_{*}$,
 where the magnetic field is entirely toroidal.
 The field lines about this center are given by

\begin{equation}
\label{eqno17}
{B_{r}\over dr}\,=\,{B_{\theta}\over rd\theta}\,
 =\,{B_{\phi}\over r\sin\theta d\phi}\,\,\,.
\end{equation}

\noindent By axisymmetry, the third group
 is decoupled from the first two groups.
 For the field lines on an $(r-\theta)$ plane,
 we consider the first equality between $B_{r}$ and $B_{\theta}$
 which gives $P=R(r)\Theta(x)$ equals to a constant or

\begin{equation}
\label{eqno18}
P(r,x)\,=\,R(r)\Theta(x)\,=\,C\,\,\,.
\end{equation}

\noindent In other words, the nested poloidal field lines
 are given by the contours of $P(r,x)$ on the $(r-x)$ plane.
 At the topological center, we have $\Theta(x)$ maximum and
 $R(r)$ maximum, so that $P(r,x)$ is maximum.
 Since $r\sin\theta$ is the distance
 of a point on the $(r-x)$ plane to the z axis,
 Eq.~\ref{eqno16c} states that the line integral
 of $B_{\phi}$ around the circle on the azimuthal plane
 is measured by $2\pi Q$,

\begin{eqnarray}
\nonumber
2\pi r\sin\theta B_{\phi}\,=\,2\pi Q\,=\,\mu I_{z}\,\,\,,
\end{eqnarray}

\noindent and the center has the maximum
 of this line integral about the axis of symmetry.
 Also, it is evident that $Q$ is equivalent to the axial current,
 where the constant part $Q=Q_{0}$ amounts to a uniform current.
 As for $P$, we evaluate the poloidal magnetic flux
 by integrating Eq.~\ref{eqno16b} on the $x=0$ plane
 over a cross section to give
 
\begin{eqnarray}
\nonumber
\int_{r_{*}}^{r} 2\pi r B_{\theta}dr\,
 =\,-2\pi (P(z)-P(z_{*}))\,\,\,.
\end{eqnarray}

\newpage
\section{Near Nonlinear Force-Free High $\beta$ Equilibria}

For high $\beta$ plasmas, we take source functions

\begin{mathletters}
\begin{eqnarray}
\label{eqno19a}
Q^2(P)\,=\,a^2P^2+Q^{2}_{0}\,\,\,,
\\
\label{eqno19b}
p(P)\,=\,p_{0}P-C_{0}\,\,\,.
\end{eqnarray}
\end{mathletters}

\noindent The Grad-Shafranov equation, Eq.~\ref{eqno7}, reads

\begin{eqnarray}
\label{eqno20}
\{r^{2}{\partial^2 P\over\partial r^2}
 +\sin\theta{\partial\over\partial\theta}
 ({1\over\sin\theta}{\partial P\over\partial\theta})
 +(ar)^{2}P\}
 +\mu p_{0} r^{4}\sin^2\theta\,
 =\,0\,\,\,.
\end{eqnarray}

\noindent The first three terms correspond to
 the force-free magnetic field as can be verified
 by setting $\mu\vec J=a\vec B$ between Eq.~\ref{eqno3}
 and Eq.~\ref{eqno4}, and the last term is the plasma
 pressure balance term. Separating the variables gives

\begin{mathletters} 
\begin{eqnarray}
\label{eqno21a}
(1-x^2){d^2\Theta(x)\over dx^2}
 +n(n+1)\Theta(x)\,=\,0\,\,\,,
\\
\label{eqno21b}
r^{2}{d^2R\over dr^2}
 +[(ar)^{2}-n(n+1)]R\,
 =\,-(\frac{\mu p_{0}}{a^{4}})(ar)^{4}
 {(1-x^2)\over\Theta}\,\,\,.
\end{eqnarray}
\end{mathletters}

\noindent The first equation gives

\begin{equation}
\label{eqno22}
\Theta(x)\,=\,(1-x^2){dP_{n}(x)\over dx}\,
 =\,(1-x^2)\,\,\,,
\end{equation}

\noindent where $n=1$ is again chosen.
 This solution gives $\Theta(x)$ positive for any $x$,
 which is appropriate for small aspect ratio plasmas.
 As for the second equation, having $n=1$,
 the solution is given by $R(r)=R_{0}(r)+R_{1}(r)$,
 where $R_{0}(r)$ and $R_{1}(r)$
 are the homogeneous and particular solutions.
 The homogeneous solution is described by

\begin{equation}
\label{eqno23}
R_{0}(r)\,=\,A_{0}arj_{n}(ar)\,
 =\,A_{0}zj_{n}(z)\,\,\,,
\end{equation}

\noindent where $j_{n}(z)$ is the spherical Bessel function.
 As for the particular solution,
 denoting $A_{1}=(\mu p_{0}/a^4)$, we have

\begin{eqnarray}
\nonumber
z^{2}{d^2R_{1}\over dz^2}+[z^{2}-n(n+1)]R_{1}\,
 =\,-A_{1}z^{4}\,\,\,,
\\
\label{eqno24}
R_{1}(r)\,=\,-A_{1}(ar)^{2}\,=\,-A_{1}z^{2}\,\,\,.
\end{eqnarray}

\noindent We note that the homogeneous solutions,
 $R_{0}(r)$ and $\Theta(x)$,
 correspond to the linear or nonlinear force-free solutions
 of the first three terms of Eq.~\ref{eqno20}
 with null or finite $Q^{2}_{0}$ respectively.
 The last term for plasma pressure
 appears only in the particular solution, $R_{1}(r)$,
 that keeps the pressure balance.
 The homogeneous radial solution
 is an oscillating function in $z=ar$,
 which has sucessive maxima,
 and the homogeneous meridian solution
 has a lobe peaked at $x=0$,
 like the low $\beta$ case.
 The superposition of the particular radial solution
 only slightly modifies the homogeneous solutions.
 We could use the region between $z=0$
 and the first root of $j_{n}(z)$, with $n=1$,
 to describe low aspect ratio high $\beta$
 toroidal plasma equilibria.
 Since Eq.~\ref{eqno22}, Eq.~\ref{eqno23}, and Eq.~\ref{eqno24}
 are exact solutions of Eq.~\ref{eqno20},
 with source functions of Eq.~\ref{eqno19a} and Eq.~\ref{eqno19b},
 these exact solutions decribe high $\beta$ plasmas,
 that are nearly force-free.
 It is interesting to note that exact near force-free solutions
 appear in high $\beta$ equilibria,
 while the low $\beta$ equilibria are not force-free.

\newpage
\section{Low Aspect Ratio Fusion Tori}

These benchmark analytic equilibria are most
 relevant for current international collaboration
 scale fusion tokamaks, such as JET and ITER,
 with D shaped plasma cross section.
 The homogeneous radial force-free solution $R_{0}(r)$
 is shown in Fig.1 with $A_{0}=0.05\,Tm^2$ taken arbitrarily.
 This $R_{0}(r)$ solution has the first null point at $z_{0}=4.5$.
 To consider the particular solution $R_{1}(r)$,
 we first have to determine the constant
 $A_{1}=(\mu p_{0}/a^4)$.
 To fix the parameter $a$, we remark that the first
 stationary point, for the magnetic axis, is around

\begin{eqnarray}
\nonumber
z_{*}\,=\,ar_{*}=2.7\,\,\,.
\end{eqnarray}

\noindent Taking the major radius $r_{*}=0.5\,m$
 for a given low aspect ratio spherical tokamak,
 we would have $a=z_{*}/r_{*}=5.4/m$.
 Considering plasma pressure of $10^{5}\,Pa$, we have
 $A_{1}=1.3\times 10^{-9}p_{0}=1.3\times 10^{-4}$.
 With the given $A_{1}$, the particular solution
 of Eq.~\ref{eqno24} is shown in Fig.2.
 Superimposing the two solutions gives the near force-free
 general solution $R(z)=R_{0}(z)+R_{1}(z)$
 which shifts the null point of $R(z)$ to $z_{0}<4.5$,
 where the plasma pressure vanishes.
 The corresponding contour point of $z_{0}<4.5$
 on the inner side of the torus is $z_{0}=0$,
 where the plasma pressure also vanishes.
 However, there are engineering aspects of fusion torus,
 such as divertor scrape-off effect,
 that alter the outer and inner $z_{0}$ plasma boundaries.
 This effectively means that the plasma boundary
 is set by the constant $C_{0}$
 in Eq.~\ref{eqno19b} or Eq.~\ref{eqno10b}
 determined by engineering considerations.

The poloidal magnetic contours of Eq.~\ref{eqno18}
 are shown in Fig.3.
 Since the plasma pressure of Eq.~\ref{eqno19b}
 is expressed in terms of $P$,
 it shares the same D shaped contours of Fig.3.
 As for the toroidal field, with $Q_{0}^{2}=0$,
 the toroidal field circulation about the z axis
 is expressed in terms of $2\pi Q=2\pi aP$,
 and therefore it also shares the contours of Fig.3.
 As for the current density of Eq.~\ref{eqno4},
 making use of the Grad-Shafranov equation
 of Eq.~\ref{eqno20} gives
  
\begin{eqnarray}
\nonumber
\mu\vec J\,
 =\,{1\over r\sin\theta}
 \{+{1\over r}{\partial Q\over\partial\theta},
   -{\partial Q\over\partial r},
   +(a^2R_{0}+a^2R_{1}+\mu p_{0}r^2)\Theta\}\,
\\
\label{eqno25}
 =\,{1\over r\sin\theta}
 \{+{1\over r}{\partial Q\over\partial\theta},
   -{\partial Q\over\partial r},
   +a^2R_{0}\Theta\}\,\,\,.
\end{eqnarray}

\noindent The middle equality is obtained by cancelling
 the $R_{1}$ term with the $\mu p_{0}$ term
 using Eq.~\ref{eqno24}.
 Analogous to the magnetic field lines,
 the current density field lines are given by
 
\begin{equation}
\label{eqno26}
{J_{r}\over dr}\,=\,{J_{\theta}\over rd\theta}\,
 =\,{J_{\phi}\over r\sin\theta d\phi}\,\,\,.
\end{equation}

\noindent Considering the first equality,
 the poloidal current density contours are given by

\begin{eqnarray}
\label{eqno27}
Q(r,x)\,=\,C\,\,\,.
\end{eqnarray}

\noindent As a result of this,
 Fig.3 also represents the poloidal current density contours
 in the linear force-free case of $Q^{2}_{0}=0$
 such that $Q(P)=aP(r,x)=aR(r)\Theta(x)$.
 
The $B_{\phi}$ field along $x=\cos\theta=0$ in Tesla,
 Eq.~\ref{eqno16c}, is shown in Fig.4.
 Since this field scales as $R(r)/r$,
 it has a $j_{1}(r)$ profile of Fig.1,
 and it peaks around $z=2.1$.
 The $B_{\theta}$ field along $x=\cos\theta=0$ in Tesla,
 Eq.~\ref{eqno16b}, is also shown in Fig.4.
 This component scales as $(1/r)(dR(r)/dr)$,
 and it crosses zero around $z=z_{*}=2.6$
 which sets the magnetic axis.
 We note that the magnetic axis at $z=2.6$,
 where $B_{\theta}$ vanishes,
 and the stationary point of the toroidal field at $z=2.1$,
 where $B_{\phi}$ is maximum, do not coincide.
 They are separated by
 $\Delta z=a\Delta r=0.5$, or $\Delta r=0.1\,m$.
 Since the toroidal field contours are off-centered
 from the poloidal field contours,
 the magnetic pressure contours are not aligned
 with the plasma pressure contours.
 The discrepancy is due to the magnetic curvature
 which gives rise to magnetic tension according to
 $\mu\vec J\times \vec B=(\vec B\cdot\nabla)\vec B-\nabla B^2/2$.
 This difference $\Delta z$ would be reduced
 if we use subsequent second and third peaks of $R_{0}(r)=6.1,9.3$
 and corresponding $j_{1}(r)=6.0,9.2$, not shown in Fig.1.
 As an example, for the second peak, we would have $z_{*}=6.1$
 and $\Delta z=0.1$. With major radius $r_{*}=2\,m$,
 this would give $\Delta r=(\Delta z/z_{*})r_{*}=0.03\,m$,
 and would have high aspect ratio equilibria.
 The toroidal and poloidal current densities,
 $J_{\phi}$ and $J_{\theta}$,
 along $x=\cos\theta=0$ in $10^6\,A/m^2$ are shown in Fig.5.
 Besides the scale difference, Fig.4 and Fig.5
 are slightly different with $J_{\phi}/B_{\phi}$
 slightly larger than $J_{\theta}/B_{\theta}$.
 Had not been the plasma pressure dependent
 particular solution $R_{1}(z)$,
 they would be the same,
 because of the force-free nature
 with $\vec J$ and $\vec B$ being proportional.
 
\newpage
\section{Safety Factor}

One important parameter that measures the quality
 of equilibrium is the safety factor,
 which measures the number of turns
 that the toroidal field circles
 about the toroidal axis of symmetry
 through the major radius
 as the poloidal field makes one turn
 about the magnetic axis
 through the minor radius
 returning to the same poloidal angle,
 but not necessarily the same toroidal angle.
 In spherical coordinates, 
 this amounts to the winding number
 of the toroidal field about the z axis of symmetry
 as the poloidal field circles one complete turn
 about the magnetic axis.
 With our spherical coordinates,
 the poloidal field is composed of $B_{r}$ and $B_{\theta}$,
 and the usual safety factor expression does not apply.
 In order to rederive an expression for the safety factor,
 we consider the field line equation, Eq.~\ref{eqno17}.
 Since $B_{r}$ and $B_{\theta}$ are constrained
 by the poloidal field line contours,
 we need only one of them to track $B_{\phi}$ around the torus.
 By considering the radial component,
 and writing $Q^{2}_{0}=a^{2}\gamma^{2}$,
 such that $Q(P)=a(P^{2}+\gamma^{2})^{1/2}$,
 the field line around the torus is given by

\begin{eqnarray}
\nonumber
2\pi q(z_{1})\,=\,2\pi q(z_{2})\,=\,\int d\phi \,
 =\,\oint \frac{B_{\phi}}{B_{r}}
 \frac{dr}{r\sin\theta}\,
 =\,\oint \frac{Q(P)}{\partial P/\partial\theta}
 \frac{dr}{r\sin\theta}\,
\\
\nonumber
 =\,\frac{1}{2}\oint
 \frac{[R^2(z)(1-x^2(z))^2+\gamma^2]^{1/2}}{R(z)(1-x^2(z))}
 \frac{dz}{x(z)}\,
\\
\label{eqno28}
 =\,\oint^{z_{2}}_{z_{1}}
 \frac{[R^2(z)(1-x^2(z))^2+\gamma^2]^{1/2}}{R(z)(1-x^2(z))}
 \frac{dz}{x(z)}\,\,\,,
\end{eqnarray}

\noindent where the integration is carried out
 over the contour $R(z)\Theta(x)=C$, as in Fig.3,
 with $x(z)$ in the integrant given by

\begin{equation}
\label{eqno29}
x(z)^2\,=\,(1-\frac{C}{R(z)})\,
 =\,\frac{1}{R(z)}(R(z)-C)\,\,\,.
\end{equation}

\noindent The integration is bounded by $z_{1}<z<z_{2}$
 on the $x=0$ horizontal plane,
 where $x_{1}=x(z_{1})=0$ and $x_{2}=x(z_{2})=0$
 are the first and second intercepts
 of a given contour on the horizontal axis.
 Although the integrant is singular
 at the bounds $z_{1}$ and $z_{2}$, it is integrable.
 As the contour value $C$ increases towards the maximum of $R(z)$,
 the contour closes onto the magnetic axis of $z=z_{*}$,
 and $x_{1}$ and $x_{2}$ come closer together towards $z_{*}$.
 In this limit, $x(z)$ in the denominator approaches zero,
 and the integration interval $z_{1}$ and $z_{2}$
 of Eq.~\ref{eqno28} also reduces to null.
 As a result, the contour integral near the magnetic axis
 depends on the limiting integral of the last factor $dz/x(z)$.
 To get the correct limit,
 we Taylor expand $R(z)$ about $z_{*}$ to second order to obtain
 
\begin{eqnarray}
\label{eqno30}
2\pi q(z_{*})\,
 =\,\frac{[R^2(z_{*})+\gamma^2]^{1/2}}{R(z_{*})}
 \oint^{z_{2}}_{z_{1}}\frac{dz}{x(z)}\,
 =\,\{\frac{2[R^2(z_{*})+\gamma^2]}
 {R(z_{*})|\partial^{2}R(z_{*})/\partial z^{2}|}\}^{1/2}
 \ln\frac{|z_{2}-z_{*}|}{|z_{1}-z_{*}|}
 \,\,\,.
\end{eqnarray}
  
\noindent Since the $R(z)$ profile is not symmetric around $z_{*}$,
 as shown in Fig.1,
 $z_{1}$ and $z_{2}$ of a given contour value
 do not approach $z_{*}$ at equal intervals,
 and $2\pi q(z_{*})$ is finite.
 Since the second order term of the Taylor expansion
 is a symmetric quadratic term in $(z-z_{*})$,
 this left-right non-symmetry of the maximum
 comes from the third order cubic term.

As an example, let us consider $\gamma^2=0.01$.
 Since the maximum value of $P=R\Theta$ is about $0.05$,
 according to Fig.1 with $\Theta=1$,
 the constant $\gamma^2$ term dominates the $P^2$ term in $Q$.
 By Eq.~\ref{eqno19a}, this amounts to an almost uniform
 axial current $\mu I_{z}=2\pi Q$.
 In this case, the scalar $K(P)={\partial Q/\partial P}$
 is not a constant, but a function of $P$,
 which makes the force-free equation nonlinear.
 The difference between the linear and nonlinear force-free fields
 is that the toroidal magnetic field $B_{\phi}$
 vanishes at some radial positions for linear case,
 while it is always poisitive definite for nonlinear case.
 With the corresponding parameters in Figs.1-5,
 the safety factor profile is shown in Fig.6,
 which ressembles well with laboratory profiles,
 having a basin within the plasma interior
 surrounded by high values on the border.
 The value at the border can be lowered
 by reducing the constant $\gamma^2$.
  
\newpage
\section{Conclusions}

We have examined the axisymmetric toroidal
 plasma equilibria for low and high $\beta$ plasmas.
 The center of the spherical system, $z=0$, has been excluded
 because this point usually lies outside the plasma domain
 due to the machine vessel.
 However, for spheromaks, the center is connected to the plasma domain.
 In this case, our solutions provide hollow profiles for plasma pressure.
 Rather than solving the Grad-Shafranov equation
 in toroidal geometry with cylindrical coordinates,
 devised to envelope axisymmetric toroidal plasmas,
 we have taken a different approach to represent
 the Grad-Shafranov equation in spherical coordinates
 and have solved for toroidal solutions.
 This approach allows benchmark plasma equilibria,
 such as the generalized Solovev type low $\beta$ equilibrium
 and the standard high $\beta$ equilibrium,
 be solved by separation of variables
 in terms of elementary special functions,
 and provides easier visualization of the solutions.
 The essential features of the solutions
 in spherical coordinates
 are that the radial function $R(z)$
 has a maximum at $z=z_{*}\neq 0$,
 and the meridian function $\Theta(x)$
 has a lobe at $x=0$.
 These two features assure the function
 $P(z,x)=R(z)\Theta(x)$
 an axisymmetric structure of a torus.
 Realistic high $\beta$ near force-free D shaped contours
 are evident in these analytic representations.

In the generalized Solovev type source functions,
 the equilibrium is solved approximately
 by sucessive approximation to the first order.
 As a result, the validity of the solution
 requires strictly a low $\beta$ plasma.
 For more specific source functions,
 the equilibrium is solved exactly in closed form.
 The solution is, therefore, valid for high $\beta$ plasmas.
 Surprisingly, the high $\beta$ exact solution
 is constructed on a force-free magnetic field
 superimposed by a high $\beta$ plasma contribution.
 Although with high $\beta$, the plasma contribution
 is small comparing to the force-free part,
 making the overall solution near force-free.
 The reason for this near force-free solution
 to confine plasmas is that
 it relies basically on the tensional force
 of the magnetic field line curvature,
 not the scalar magnetic pressure,
 to balance the plasma pressure.
 This curvature is evident
 because the magnetic axis where $B_{\theta}=0$
 is significantly displaced from the maximum of $B_{\phi}$.
 For this reason, the magnetic field profile
 remains near force-free, while plasmas are confined.
 Such strong curvature is inherent from low aspect ratio plasmas.
 The absence of such curvature in high aspect ratio plasmas
 obliges the machines to operate in low $\beta$ limit
 with magnetic fields away from force-free configuration.


{}

\clearpage
\begin{figure}
\plotone{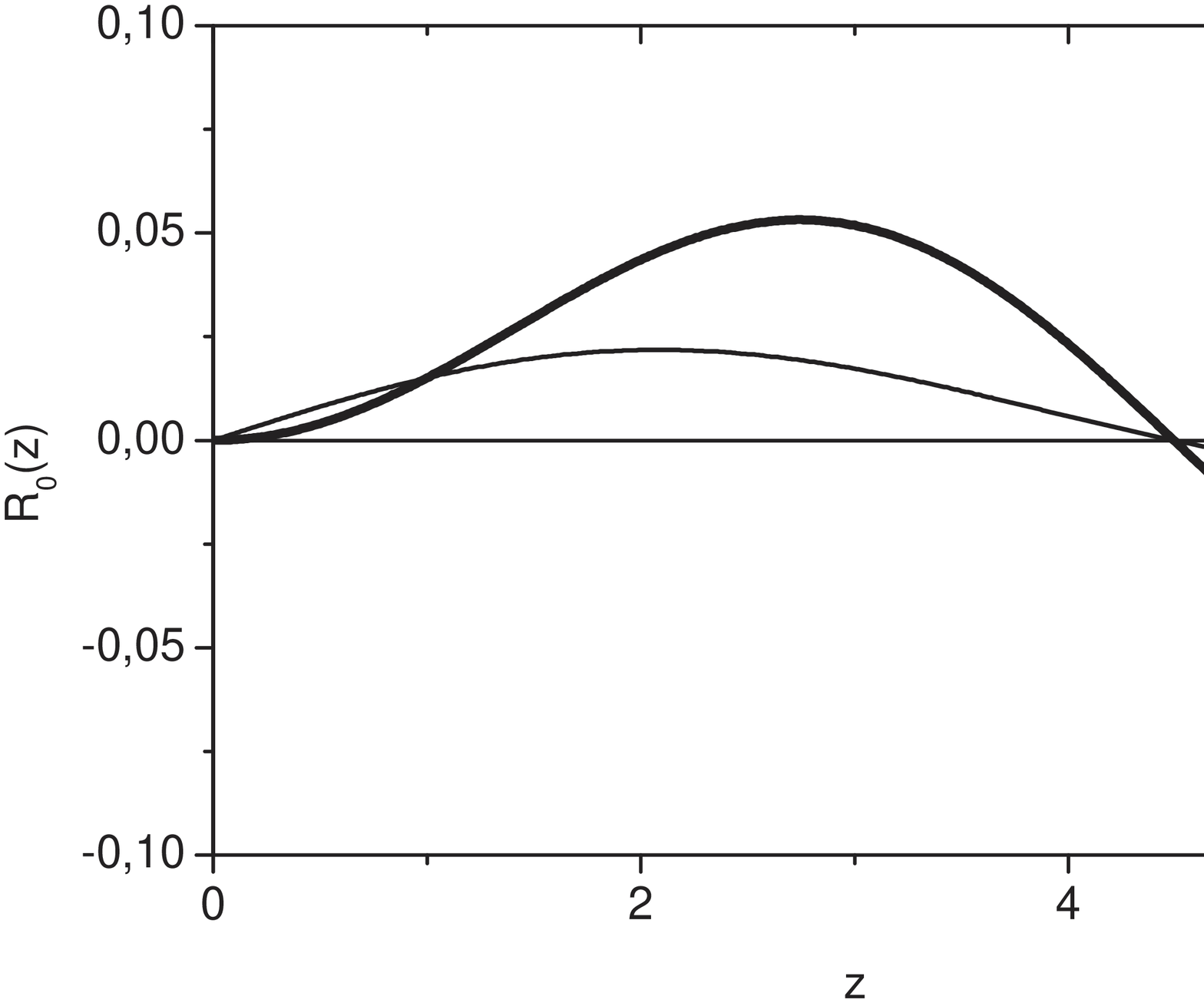}
\caption{The function $R_{0}(z)=A_{0}zj_{1}(z)$ with
 $A_{0}=0.05$ of the force-free homogeneous radial solution
 is plotted as a function of $z$ in thick line.
 The spherical Bessel function $A_{0}j_{1}(z)$ is also plotted
 in thin line for comparisons.}
\end{figure}

\clearpage
\begin{figure}
\plotone{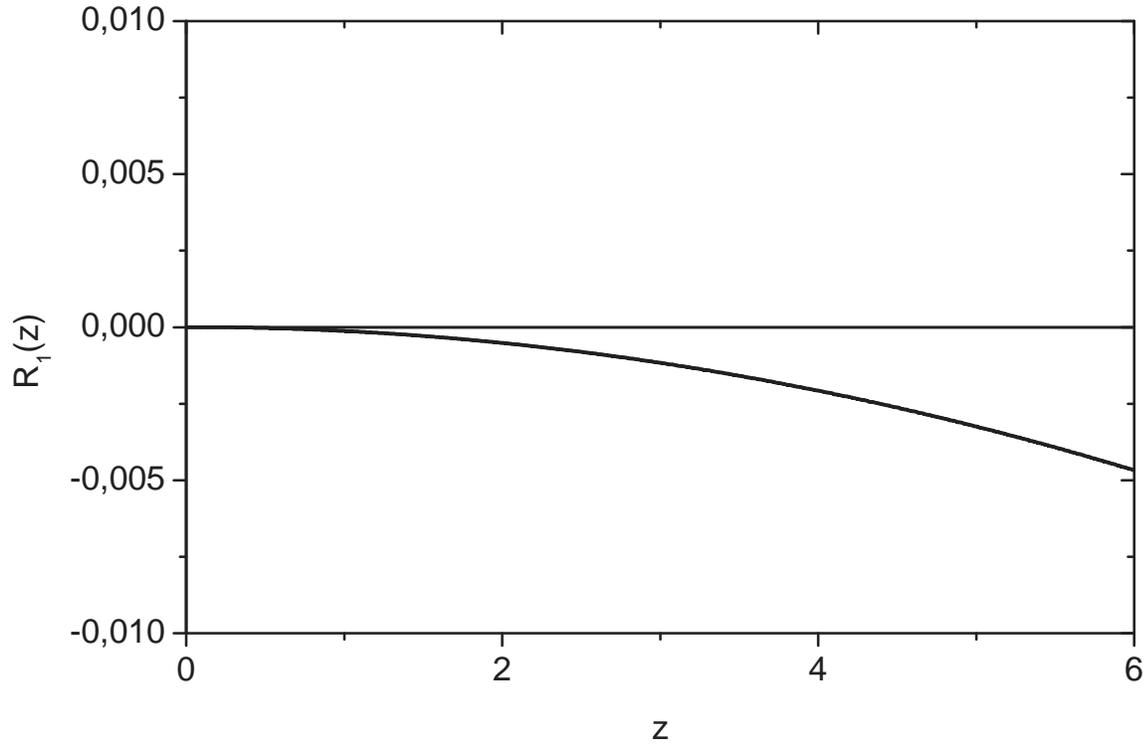}
\caption{The function $R_{1}(z)=-Az^{2}$ with
 $p_{0}=10^{5}\,Pa$ and $A=1.3\times 10^{-4}$
 of the particular radial solution
 is plotted as a function of $z$
 to show the plasma pressure effect.}
\end{figure}  

\clearpage
\begin{figure}
\plotone{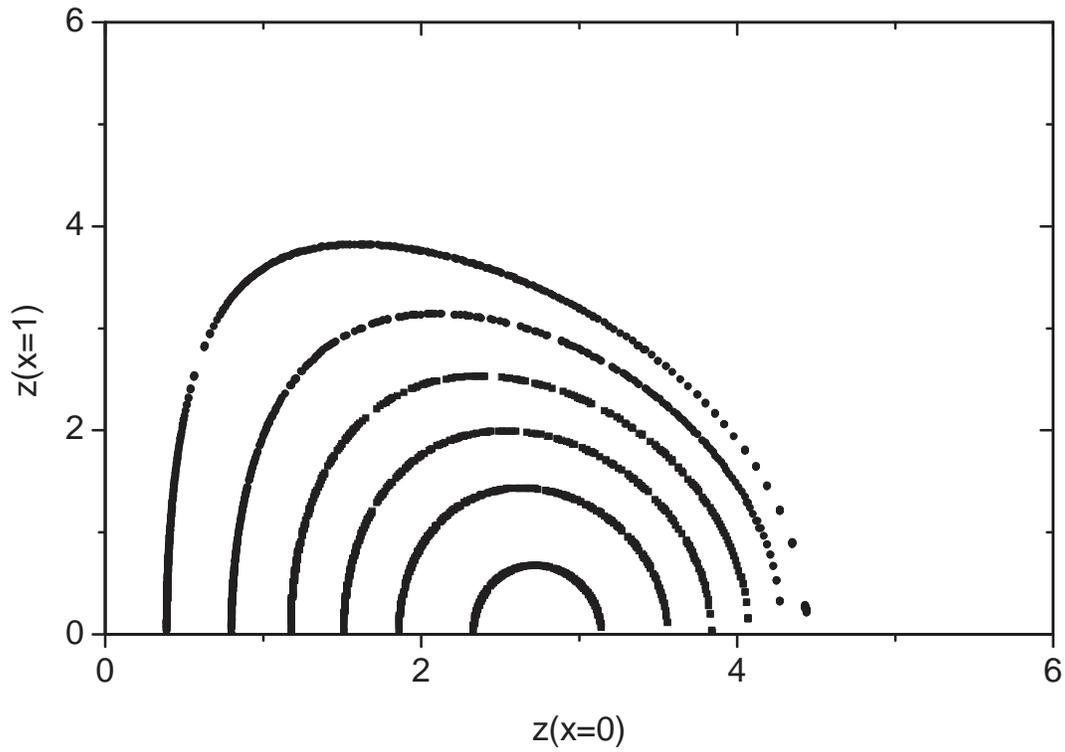}
\caption{The magnetic field lines for spherical tokamak
 are shown with increasing contour labels
 from outer to inner contours.
 These contours are shared by the plasma pressrue and
 the toroidal field line integral.
 The axes are labelled in $z$ together with $x=\cos\theta$.}
\end{figure}

\begin{figure}
\plotone{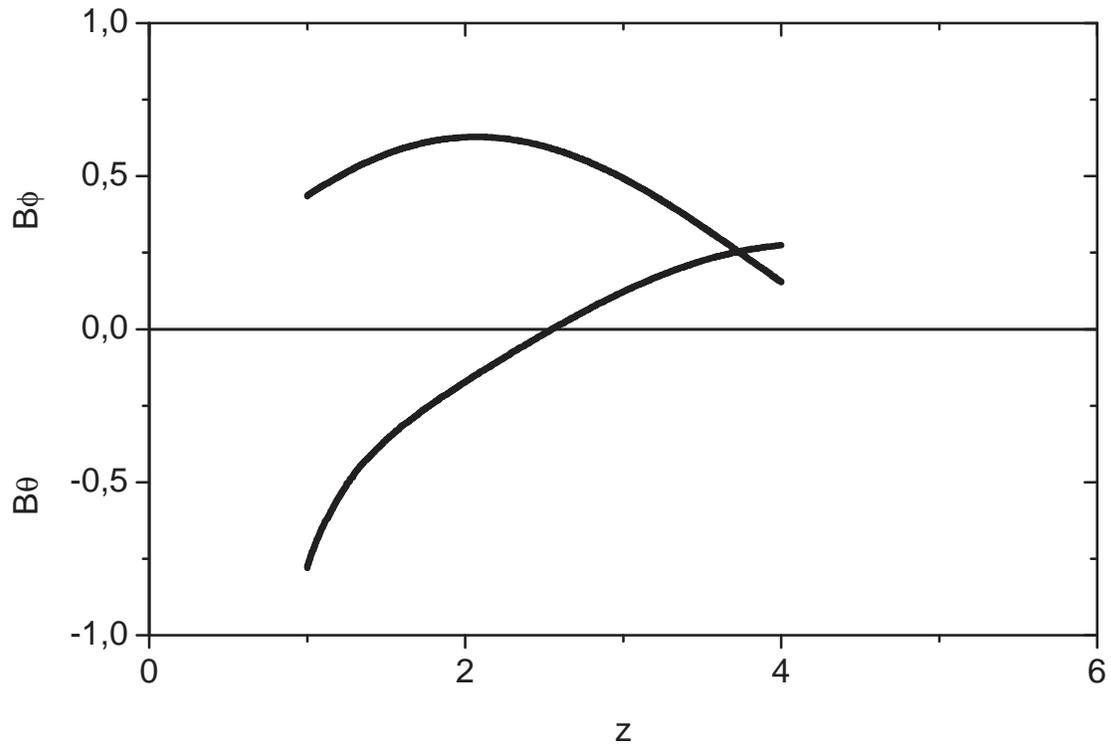}
\caption{The toroidal and the poloidal magnetic fields,
 $B_{\phi}$ and $B_{\theta}$, with $Q=aP$ and $a=5.4$
 are shown as a function of $z$ along $x=0$.}
\end{figure}

\begin{figure}
\plotone{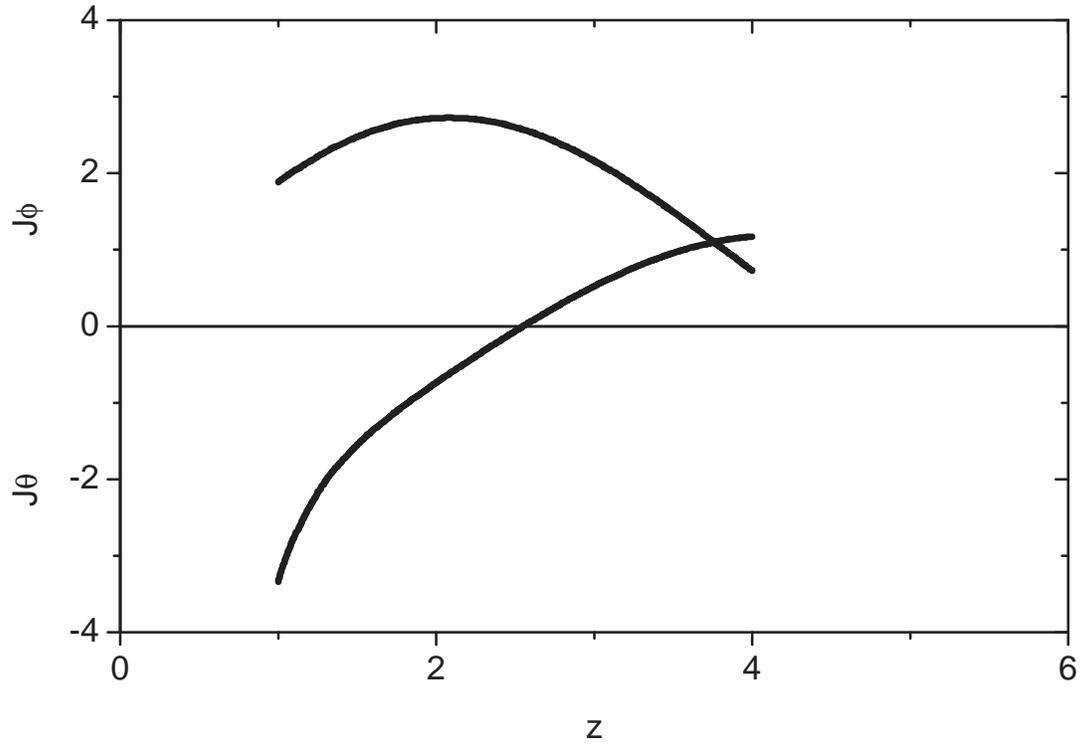}
\caption{The toroidal and the poloidal current densities,
 $J_{\phi}$ and $J_{\theta}$, with $Q=aP$ and $a=5.4$
 are shown as a function of $z$ along $x=0$
 in units of $10^6\,A/m^2$.}
\end{figure}

\begin{figure}
\plotone{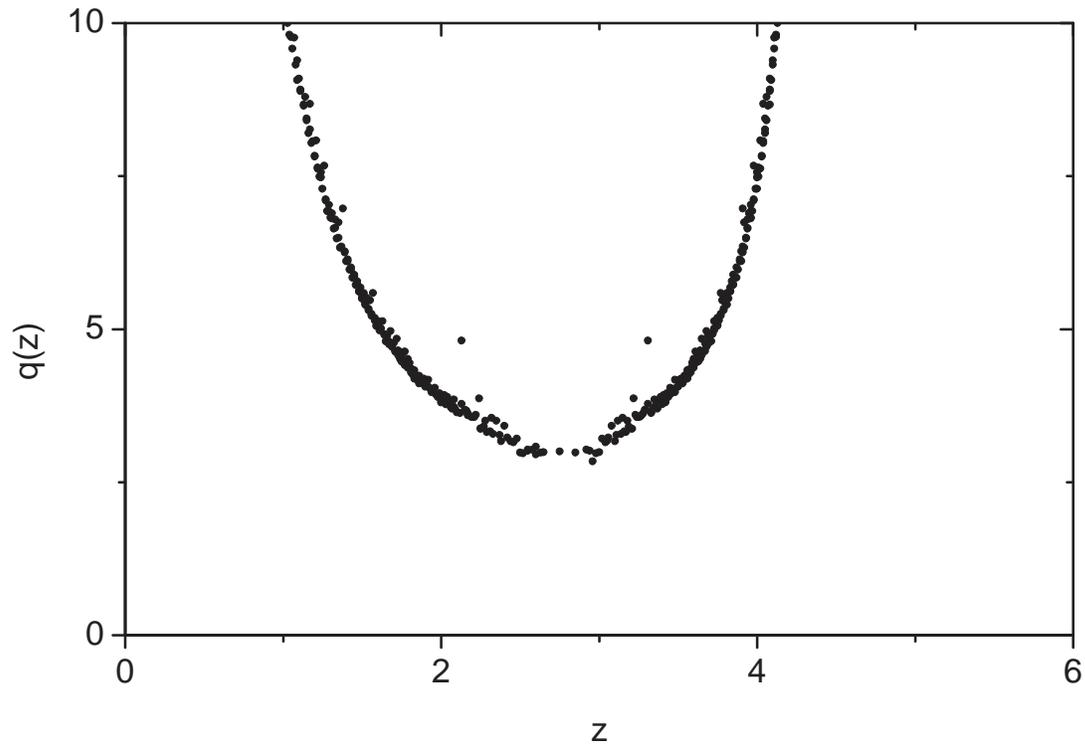}
\caption{The safety factor profile of the plasma equilibrium
 with nonlinear force-free homogeneous solution, $\gamma^2=0.01$,
 is shown as a function of $z$ along $x=0$.}
\end{figure}

\end{document}